\documentclass[letterpaper, 10 pt, conference]{ieeeconf}  

\IEEEoverridecommandlockouts                              

\usepackage{graphics} 
\usepackage{epsfig} 
\usepackage{mathptmx} 
\usepackage{times} 
\usepackage{amsmath} 
\usepackage{amssymb}  
\usepackage{algorithmicx}
\usepackage{algorithm}
\usepackage{algpseudocode}
\usepackage{amsfonts}
\usepackage{xcolor}
\usepackage{soul}
\usepackage{flushend}
\usepackage{url}

\title{\LARGE \bf
A Learning-based Optimal Market Bidding Strategy \\ for Price-Maker Energy Storage
}

\author{Mathilde D. Badoual$^{1}$ and Scott J. Moura$^{1}$
\thanks{$^{1}$M. Badoual and S. Moura are with the Energy, Controls and Applications Lab (eCAL) at University of California, Berkeley, USA.
        {\tt\small \{mathilde.badoual, smoura\}@berkeley.edu}}%
        }

\begin{document}

\maketitle
\thispagestyle{empty}
\pagestyle{empty}

\begin{abstract}

Load serving entities with storage units reach sizes and performances that can significantly impact clearing prices in electricity markets. Nevertheless, price endogeneity is rarely considered in storage bidding strategies and modeling the electricity market is a challenging task. Meanwhile, model-free reinforcement learning such as the Actor-Critic are becoming increasingly popular for designing energy system controllers. Yet implementation frequently requires lengthy, data-intense, and unsafe trial-and-error training. To fill these gaps, we implement an online Supervised Actor-Critic (SAC) algorithm, supervised with a model-based controller -- Model Predictive Control (MPC). The energy storage agent is trained with this algorithm to optimally bid while learning and adjusting to its impact on the market clearing prices. We compare the supervised Actor-Critic algorithm with the MPC algorithm as a supervisor, finding that the former reaps higher profits via learning. Our contribution, thus, is an online and safe SAC algorithm that outperforms the current model-based state-of-the-art.

\end{abstract}

\vspace{0.2cm}



\section{INTRODUCTION} \label{intro}


\subsection{Background \& Motivation}

Large-scale energy storage systems can solve a number of issues that can arise on electric power systems with high penetration of intermittent renewable energy generation.  Energy storage and more specifically, lithium-ion batteries which are particularly fast to charge and discharge, can help to keep the voltage within bounds, stabilize grid frequency, and provide reserves that can be called on should a contingency occur. Storage can also address issues related to surplus that can occur when renewable sources produce more electricity than consumers demand, and can mitigate \textit{congestion} when the transmission capacity of the system is exceeded. Because legacy grid technologies are unable to store electricity once produced, electricity prices vary significantly between times of surplus and shortage. This allows batteries to conduct \textit{arbitrage}, and gives them a competitive edge on the energy market. \cite{barton2004} \cite{carrasco2006} \cite{taylor2013}. 

In 2018, the U.S. Energy Information Administration \cite{us_report} reported that operational large-scale battery storage represented 869 megawatts (MW) of power capacity and 1,236-megawatt hours (MWh) of energy capacity. Moreover, markets across the globe are implementing new regulations that will create new value streams for large-scale batteries \cite{xu2016}. In the US, the Federal Energy Regulatory Commission issued an order in 2011 requiring transmission grid operators -- including Independent System Operators and Regional Transmission Operations (ISO/RTO) -- to compensate fast-ramping resources on the frequency regulation market \cite{kintner-meyer2014}. In Europe, the Frequency Control Regulation (FCR) market is updated to respond to the increased need for battery storage \cite{web:europe_fcr}. These new market rules favor grid-scale storage resources, which have response capabilities that conventional generation resources do not. These market incentives have led to increased investment in energy storage capacity. 

The increase in storage capacity coupled with a unique position in the market has caused grid-scale energy storage to become a driver of the market price. In economic terms, energy storage is said to be a \textit{price-maker}, or a monopolistic seller capable of influencing the market because no substitutes exist for their product. Conventional power plants, on the other hand, are said to be a \textit{price-taker} with trivial influence on the market due to market size, suitability, or other anti-competitive factors. In a perfectly competitive market, all the sellers are \textit{price-takers} and bid their marginal cost. More details on the price-maker/price-taker concepts can be found in \cite{Steeger2014} and \cite{borenstein}. 

One real-world example is the Australian Hornsdale power reserve. Upon entering the Australian market \cite{web:hornsdale1} \cite{web:hornsdale2} in December 2017, the battery achieved over 55\% of the Frequency Control Ancilliary Services (FCAS) revenues in South Australia and prices went down by 90\%. This occurred both due to the size of the battery and due to its fast-ramping capabilities that could not be matched by other market participants with pre-existing technologies.

Strategic bidding in markets with \textit{price-makers} is a very challenging problem, due to the complexity of the market model as well as the lack of information about other players. Each agent must try to anticipate not only the demand but also the actions of other market participants. However, agents generally do not have models that describe the bidding strategies of other market participants.
The current work explores the use of adaptive control for optimizing the bidding strategy of a \textit{price-maker} agent participating in a regular wholesale market.

\subsection{Literature Review}

Several papers explore optimal bidding algorithms on the electricity market when bids influence the clearing price, i.e. the market player is a \textit{price-maker}. Some relevant examples include the following: Oren et al. \cite{oren1975} computed the optimal bidding strategy with dynamic programming by estimating other market players. Kwon et al. \cite{Kwon2012} review the optimization problem for bidding in the day ahead market. Velazquez et al. \cite{Vazquez2014} base their bidding strategy on the study of the residual demand curve. 

The bidding of energy storage capacity on the electricity market adds a layer of complexity. The battery has a limited capacity and accumulates revenue by scheduling efficiently generation and load modes. J. Arteaga et al. \cite{Arteaga2019} develop a robust and stochastic optimization for the bidding on the Real-Time and Day-Ahead market in which the battery is a \textit{price-taker}. They also study participation in the ancillary services market where the storage system is \textit{price-maker} due to the battery efficiency in ramping up and down under emergency.

Throughout this literature, a common method to solve the optimal bidding strategy for a \textit{price-maker} is used. A bi-level optimization program where the first layer maximizes the player's revenue and the second layer solves a dispatch problem to maximize the social welfare. This method is developed, among others, by T\'{o}masson et al. \cite{tomasson2020} and Y. Ye et al. \cite{Ye2020}. The resulting bi-level optimization problem is complex to solve and is commonly simplified using the Karush-Kuhn-Tucker (KKT) conditions relying on the convexity of the problem which is not guaranteed. Indeed, modeling the energy market dynamics as an optimization problem yields non-linearities. For instance, the dispatch can involve binary unit commitments. Perhaps more importantly, modeling the entire market requires knowledge of other players' strategies and predictions of their future behaviors. This can be difficult or even impossible due to lack of data and highly non-predictable behavior. Moreover, market prices can be volatile and depend on physical constraints, e.g. congestion, frequency instabilities. 

A possible solution for this problem is to use learning-based and adaptive controllers, such as reinforcement Learning (RL) which learns a controller (or policy) via structured trial-and-error. Such techniques enable design of controllers for partially known systems. Moreover, the learned controller becomes a function mapping the state to an action (state-feedback control), which makes the controller computationally efficient. Glavic et al.  \cite{Glavic2017} review the literature of Reinforcement Learning in electric power system decision making, presenting recent breakthroughs in Reinforcement Learning such as ``safe RL'' and ``path integral control for RL''. More recently Z. Zhang et al. \cite{8859593} review the application of Deep Reinforcement Learning (DRL) -- Reinforcement Learning with the use of Deep Neural Networks -- in Power Systems. They point out the benefits of DRL as an adaptive and model-free controller applied on electricity market trading in the case of incomplete information. 

Among others, Gajjar et al. \cite{Gajjar2002} use an Actor-Critic Algorithm to solve the bidding problem. The authors construct an Actor-Critic algorithm to model the behavior of generating companies, formalizing a mathematical model for competitor's behavior. However the battery is considered as a \textit{price-taker} on the market. As a result, the paper does not investigate the use of DRL in an environment reacting to its action. Y. Ye et al. \cite{Ye2020} use a form of Policy Gradient method in the case of a \textit{price-maker} battery. The algorithm is a Deep Deterministic Policy Gradient with a Prioritized experience replay. In other words, Ye et al. use a policy gradient trained using a prioritized sample batch. The paper reports interesting results but does not tackle the safety issue of trial and error training on a real market, neither the complexity of building a realistic market environment for an off-line training. Indeed, the main justification for the use of RL and DRL lies in the complexity of the market model which makes the optimization of the bidding strategy a complex task. However, if this model can be built to train an adaptive algorithm, then optimization techniques will be more safe and reliable than a trial and error training.

Safety in DRL can take multiple definitions. In our context, safety is achieved when actions taken by the DRL algorithm, also called agent or policy, are not putting the system in states that are physically impossible to reach and that can lead to dangerous situations, such as discharging an empty battery. 

\subsection{Contributions}
The main contribution of this paper is to ensure safety and improve performance via an online learning-based algorithm for optimal energy storage bidding, under \textit{price-maker} conditions. Specifically, we develop a Supervised Actor-Critic algorithm. The supervisor technique reduces the action space dimension and thus accelerates the learning process while ensuring that the algorithm tries less dangerous actions. The supervisor is a naive and model-based algorithm, MPC, with the assumption that the battery is a \textit{price-taker}. This algorithm is deterministic and does not take dangerous actions. In addition, we develop a shield, to protect the battery from charging or discharging above limits. Finally, we add a penalty term to the reward function to inform the algorithm of dangerous actions.  We achieve a more efficient bidding strategy than the baseline model-based technique, while also ensuring safety during training. 

\subsection{Paper Organization}
Section II, details the market model and simulation. Section III introduces the Supervised Actor-Critic and Section IV describes the results. 

\section{Market and Storage System Model} \label{model}
In this section we describe the market setup and energy storage system operation used in our simulation. 
\begin{figure}[H]
\centering
\includegraphics[width=7cm]{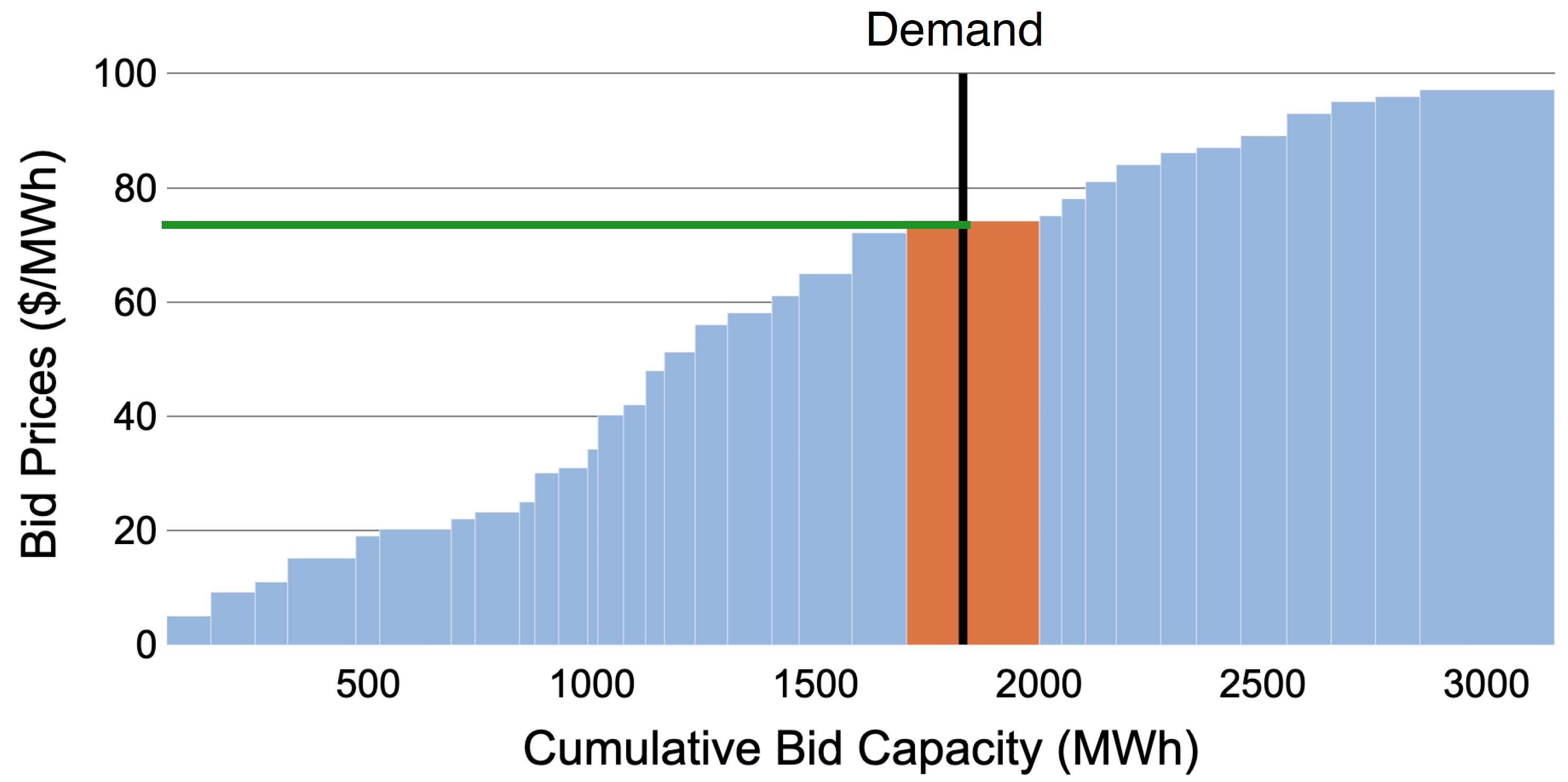}
\caption{The Market Clearing Process as a Uniform-Price Auction}
\label{clearing_process}
\end{figure}
The electricity market is managed by the Independent System Operator (ISO) and maximizes the social welfare across generators and consumers so that enough generation is dispatched to serve a predicted demand. One step of the bidding process is represented in Fig. \ref{clearing_process}, where the demand is the vertical black line. At the point where the cumulative bid capacity reaches the demand, the market clears at this bid price, represented by the green line on the figure. This price is the so-called clearing price. The market used in this paper follows a \textit{uniform-price auction} \cite{Akbari-Dibavar2020}, i.e. the actors are paid the clearing price instead of being paid at their bidding price. The clearing process can be written as the following optimization problem (1)-(3):
\begin{flalign}
    \min_{\textcolor{blue}{P^t}} \quad & \sum_{i=0}^{n} c^t_i \textcolor{blue}{P^t_i} \\
    \textrm{s.t} \quad \forall i, \quad & 0 \leq \textcolor{blue}{P^t_i} \leq p^t_{i} \\
    & \sum_{i=0}^{n} \textcolor{blue}{P^t_i} = \hat{d}_t 
\end{flalign}
Each agent is represented by the index $i$ and sends a bid at each time step $t$. A bid is of the form $\{c^t_i, p^t_i\}$ where $c^t_i$ is the cost per unit energy and $p^t_i$ is the  capacity that the agent $i$ can provide during time $t$. The variable $\hat{d}_t$ is the total predicted energy demand from retailers, industrial sites or storage systems. The market solves the social welfare maximization problem (1) by dispatching a capacity $P^t_i$ for each agent while matching supply and demand (3). This capacity is either 0 when the bid is not cleared, or a number between 0 and $p^t$, the capacity bid by the agent (2). The optimization variable is $P^t = [P^t_0 \ \cdots \ P^t_n]^T$ with $n$ the number of agents.  

The complete bidding process of the battery owner or agent can be described as follows, and represented in Fig. \ref{work_flow}. The agent sends a bid to the market. The market clears the bids depending on the demand and according to the process described in Fig. \ref{clearing_process}. Then, if the battery bid is cleared amongst all bids to the left of the black line in Fig. \ref{clearing_process}, a command is passed to the battery to provide the capacity dispatched by the market at the cleared price. The combination of the market state and the battery state is sent back to the battery's bidding agent to compute a new bid at the next step. 


Batteries generally have a larger impact on ancillary service markets and especially on frequency control markets. The reason is that, in small markets, the battery's fast response give them a competitive advantage over other players \cite{BERRADA20161109}. Here, we focus on the use and limitations of reinforcement learning as a bidding algorithm. We use a stylized representation of market dynamics in order to focus more specifically on the algorithm itself.

\section{Bidding Strategy Algorithm}

In this section we introduce two algorithms: the first is a common Model Predictive Control (MPC) algorithm that computes the optimal bidding strategy over a one day horizon. It will serve as our benchmark algorithm, and importantly it does not take into account that the battery is a \textit{price-maker} on the market. The second and novel approach is an adaptive controller which learns and adapts to the \textit{price-maker} situation. Specifically, the proposed controller is trained using a new SAC algorithm where the supervisor is the aforementioned MPC controller.

\begin{figure}[t!]
\centering
\includegraphics[width=9cm]{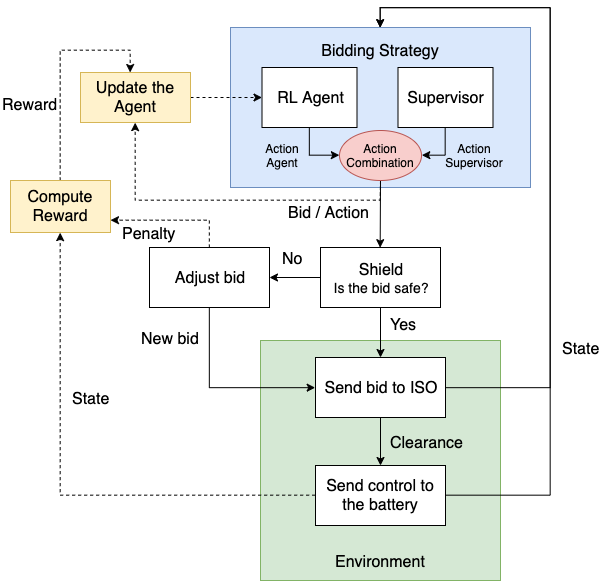}
\caption{Learning and Bidding procedure for the proposed SAC. Bids are produced by combining actions from the MPC supervisor and the DRL agent. A shield adjusts bids to ensure physical and market constraints are satisfied by allowing only feasible bids. After the market clears and the battery state evolves, the reward (i.e. profit) is used to update the DRL agent.}
\label{work_flow}
\end{figure}

The online training of the SAC during the simulation is detailed in Fig. \ref{work_flow}, and explained in the next sections. 

\subsection{Model Predictive Control as a RL Supervisor}
Under the \textit{price-taker} assumption, the energy prices are considered exogenous to the bidding strategy, and the optimal bidding strategy is computed regardless of its impact on the market price. In this case, the strategy relies on predictions of the clearing price: $\hat{\lambda}_{energy}^t$ and is formulated as a common deterministic Model Predictive Control algorithm:
\begin{flalign}
    \max_{\textcolor{blue}{p_l}, \textcolor{blue}{p_g}, \textcolor{blue}{soe}, \textcolor{blue}{m}\in \mathbb{R}^H} \quad & \sum_{t=T}^{T+H} \hat{\lambda}^t (\textcolor{blue}{p^t_g} - \textcolor{blue}{p^t_l}) \\
    \textrm{s. to:} \quad \forall t, \quad &  \textcolor{blue}{soe^{t+1}} = \textcolor{blue}{soe^{t}} - \eta_g \textcolor{blue}{p^t_g}  + \eta_l \textcolor{blue}{p^t_l} \\
    & \textcolor{blue}{m^t} \in \{0, 1\} \\
    & 0 \leq \textcolor{blue}{p^t_l} \leq \textcolor{blue}{m^t} \overline{p_l} \\
    & 0 \leq \textcolor{blue}{p^t_g} \leq (1 - \textcolor{blue}{m^t}) \overline{p_g} \\
    & \underline{soe} \leq \textcolor{blue}{soe^{t}} \leq \overline{soe}
\end{flalign}

The cost function (4) is the cumulative revenue for the energy storage system owner over the horizon $H$. The variables $p_l^t$ and $p_g^t$ are the load and generation bids with price $\hat{\lambda}^t$. 
Equation (5) describes the storage system dynamics, with coefficients $\eta_g$ and $\eta_l$ referring to the charging and discharging efficiency coefficients. 
Since the storage system agent is not allowed to participate on the energy market as a load \emph{and} a generator simultaneously, constraint (6) forces the solution to chose between one of the two products (7) and (8). The remaining constraints (9) are physical limits on the storage system capacities. Other physical constraints as temperature could be added in the form of linear constraints without changing the structure of the problem solved in this paper. 

In a perfectly competitive uniform-price auction market the agents are bidding their marginal cost. The marginal cost of the battery can be seen as the cost of electricity at the time of generation, thus the battery bids the expected clearing price $\hat{\lambda}^t$. As a \textit{price-taker} the battery will not impact the market clearing price and can be sure to clear if its bidding price is less than or equal to the market clearing price. The complete bid can be written as: $\{c^t, p^t\} = \{\hat{\lambda}^t, p_g^t - p_l^t\}$.

The problem is formulated for a horizon $H$, which allows the battery to plan its charging and discharging schedule depending on the variation of electricity prices. In MPC, only the first step of the solution bid is used and the problem is re-computed at every new time step $t$. 


\subsection{Supervised Actor-Critic Algorithm}

Reinforcement Learning (RL) theory is based on the idea that we learn how to perform an \textit{action} in an \textit{environment} by structured try-fail cycles. In other words, after performing an action, we receive a \textit{reward}, which measures the value of our action. We then learn how to increase that reward with subsequent actions. Sutton and Barto provide an excellent conceptual exposition of RL theory and application \cite{sutton2018}. 

The system at time $t$ is described by state $s^t$, and the agent's action is denoted by $a^t$. We define a reward function that quantifies the agent's performance, $r(s^t, a^t)$. The objective is to refine a policy $\pi_{\theta}$ that maximizes the expected cumulative reward of the trajectory $\tau = ((s^1, a^1), ..., (s^t, a^t))$:
\begin{equation}
J(\tau) = \sum_{t=1}^{\infty} \gamma^t r(s^t, a^t)
\end{equation}
where $\gamma$ represents a ``discount factor'' that priorities near-term gains over future and less certain events. 
We also introduce the value function: 
\begin{equation}
V(s^t) = \mathbb{E}_{a \sim \pi_{\theta}}[\sum_{i=t}^{\infty} \gamma^{i} r(a^{i}, s^{i})]
\end{equation}
The value function $V(s^t)$ represents the expected cumulative reward, starting from state $s^t$ until $t\rightarrow \infty$, under policy $\pi$. The value function satisfies  Bellman's Equation, where $s^{t+1}$ is the state at $t+1$ time step: 
\begin{equation}
V(s^t) = \mathbb{E}_{a^t \sim \pi_{\theta}}[r(a^t, s^t) + \gamma V(s^{t+1})]
\end{equation}
The Actor-Critic algorithm estimates the value function from data. To compute these estimates, we utilize Bellman's equation as a regression model. Namely, we define the temporal difference $\delta$ \cite{td} as:
\begin{equation}
\delta^t = r(a^t, s^t) + \gamma V(s^{t+1}) - V(s^t)
\end{equation}
The Actor-Critic algorithm involves an ``actor'' that runs the policy $\pi_{\theta}$ with parameters $\theta$, and the ``Critic'' which is an estimate of the value function $V_{\omega}$ with parameters $\omega$. 

The Supervised Actor-Critic algorithm is similar to AC, with the addition of a ``supervisor'' policy that also proposes an action. The final action applied is a combination of the supervisor's and the Actor's action. Introduced by Rosenstein and Barto \cite{sac}, the SAC is implemented as follows:

\textbf{Initialization:} We initialize the parameters of the Actor and the Critic, respectively  $\theta$ and $\omega$.

The Actor's parameters are first updated with the difference between the policy action $a_A^t$ and the supervisor action $a_S^t$ by one step of stochastic gradient descent on the following optimization problem:
\begin{equation}
\min_{\theta} \left\{ F_a(a_A^t, a_S^t) = \frac{1}{2} \|a_A^t - a_S^t\|^2 \right\}
\end{equation}
Then the combined action is chosen using the risk parameter $k$, where $a_E^t \sim \mathcal{N}(0,\sigma)$ is an exploration term:
\begin{equation}
    a^t = (1-k)(a_A^t+a_E^t) + k a_S^t
\end{equation}

\textbf{Critic and Actor Updates:} The Critic and the Actor's parameters are then updated using Temporal Difference and Bellman's Principle of Optimality equation with the updated value function. To update the critic, we again run one step of stochastic gradient descent on the following optimization problems:
\begin{equation}
\min_{\omega} \left\{ F_c(\delta^t) = \frac{1}{2} \|\delta^t\|^2 \right\}
\end{equation}

Finally, we update the parameters of the Actor in order to increase the value function. The update is based on Bellman's principle of optimality: 
\begin{equation}
    \theta \leftarrow \arg \max_{\theta} \mathbb{E}_{a^t \sim \pi_{\theta}}[r(a^t, s^t) + \gamma V(s^{t+1})]
\label{bellman_po}
\end{equation}

In practice (\ref{bellman_po}) is computed using the Policy Gradient Theorem with ``baseline'', detailed by Sutton and Barto \cite{sutton2018}. The ``baseline'' here is an estimate of the value function given by the Critic. The update is the following:
\begin{equation}
\theta \leftarrow \theta + \beta_2 \delta^t \nabla_{\theta} \ln{\pi_{\theta}(s^t)}
\end{equation}

Algorithm 1 details the implementation of the Supervised Actor-Critic.
\begin{algorithm}[t]
\caption{Supervised Actor-Critic}
\begin{algorithmic}[]
\State \textbf{Inputs}
\State \hspace{10pt}
$V_{\omega}(s)$, Critic value function parameterized by $\omega$
\State \hspace{10pt}$\pi^A_{\theta}$, Actor policy parameterized by $\theta$
\State \hspace{10pt}$\pi^S$, supervisor policy
\State \hspace{10pt}$\sigma$, exploration factor
\State \hspace{10pt}$\alpha$, Critic step size
\State \hspace{10pt}$\beta_1$ and $\beta_2$, Actor step size
\State \hspace{10pt}$\gamma \in [0,1]$, discount factor

\State \textbf{initialize} $\theta, \omega$ randomly
\State \textbf{initialize} $s \gets$ starting state

\State \textbf{repeat} for each time step, $t$
\State \hspace{10pt} $a^A\gets$ action given by $\pi^A_{\theta}(s)$
\State \hspace{10pt} $a^S\gets$ action given by a supervisor policy $\pi^S(s)$
\State \hspace{10pt} $\theta \gets \theta + \beta_1 \nabla_{\theta} F_a(a^A, a^S)$
\State \hspace{10pt} $a^E \sim \mathcal{N}(0,\sigma)$
\State \hspace{10pt} $a\gets (1-k)(a^A+a^E) + ka^S$
\If {$a$ is not safe}
    \State \hspace{10pt} $a\gets a^S$
\EndIf
\State \hspace{10pt} \textbf{take} action $a$, \textbf{observe} reward, $r$, and next state, $s'$
\State \hspace{10pt} $\delta \gets r + \gamma V(s') - V(s)$
\State \hspace{10pt} $\theta \gets \theta + \beta_2 \delta^t \nabla_{\theta} \ln{\pi_{\theta}(s^t)}$
\State \hspace{10pt} $\omega \gets \omega + \alpha \nabla_{\omega} F_c(\delta)$
\State \hspace{10pt} $s \gets s'$
\State \hspace{10pt} Update $k \in [0, 1]$
\State \textbf{until} terminal state reached
\end{algorithmic}
\end{algorithm}

The SAC algorithm allows the bidding algorithm to train online while satisfying physical and market constraints during exploration. When the RL agent yields higher rewards than MPC, then the coefficient $k$ can be changed from $k=0$ toward $k=0.5$ to progressively weight in favor of the RL agent's bid.

To implement this algorithm within our environment, we specify the state, the action, and the reward formulation as follows.

The \textbf{environment} or system (i.e. the storage system and market) has the following state: $t_{day}$ is the time step of the day, $\lambda$ is a price either estimated, $\hat{\lambda}$, or exactly known, ${\lambda}$, and $\hat{d}_{up}$ is the estimated demand for the market. In math, $
s^t = \left[\hat{\lambda^{t}}, \lambda^{t-1}, soe^{t}, t_{day}, \hat{d}^t \right]$.

The \textbf{action} is composed of the energy market bid at time $t$. A bid has the following structure: $a^t = \{\lambda^{t}, p_g^t - p_l^t\}$.


In order to propose safe actions (actions that lead to a feasible state) while learning, we transform our reward function using a Lagrangian with predefined Lagrange multipliers. This is equivalent to adding penalty terms to the reward. The \textbf{reward} function is as follows:
\begin{equation}
\begin{split}\label{eqn:lagrangian}
    r(a^t, s^t) = & \hat{\lambda}^t(p^t_g - p^t_l) + \mu^T \mathcal{Y}(a^t, s^t)
\end{split}
\end{equation}
where $\mathcal{Y}(a^t, s^t)$ refers to constraints (5) - (9) listed in the MPC formulation, and $\mu$ serves as weights to penalize constraints violations.

\section{Experiment and Results}

\subsection{Experimental Settings}


We run the simulation using electricity demand data from the Australian electricity market, also known as AEMO (Australian Energy Market Operator) \cite{nemweb}. The simulation is run over 5 months: June to October 2018. The other players are simulated as deterministic agents bidding a fixed quantity at a fixed marginal cost. 

\begin{figure}[H]
\centering
\includegraphics[width=8cm]{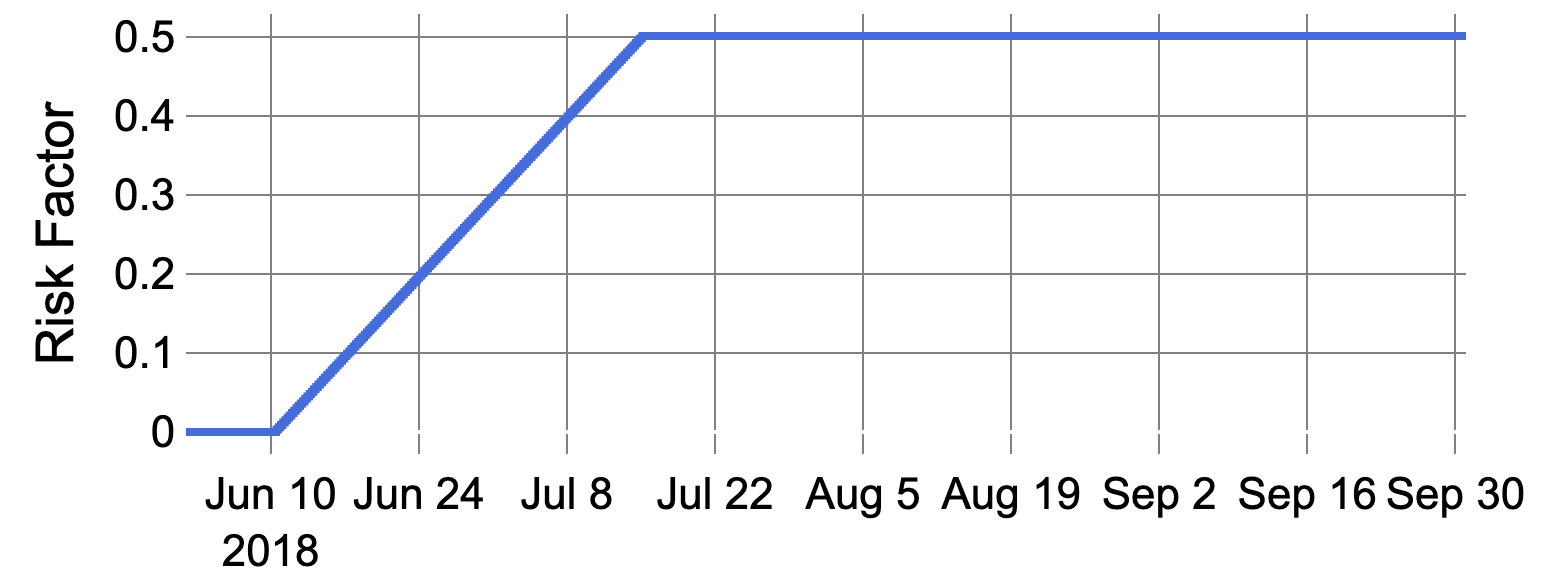}
\caption{Risk Factor Update}
\label{risk_factor}
\end{figure}

The parameters used for the supervised Actor-Critic are:
\begin{table}[!ht]
\renewcommand{\arraystretch}{1}
\caption{SAC Parameters}
\label{tab:dqn}
\centering
\begin{tabular}{c|c}
    \hline \hline
    Parameters  &  Values\\
    \hline
    \hline
    $\gamma$ discount factor & 0.98 \\
    \hline
    optimizer & Adagrad \cite{adagrad} \\
    \hline
    learning rate & $10^{-4}$ \\
    \hline
    hidden layers & 6 \\
    \hline
    Activation function & Leaky Relu \\
    \hline
    exploration $\sigma$ & 1 \\
    \hline
\end{tabular}
\end{table}
We set $k=0$ for the first 400 steps in order to select the supervisor's action while the Actor-Critic agent is training. We then progressively increase the value of $k$ until it reaches 0.5. The update of $k$ is shown in Fig. \ref{risk_factor}.
The battery is designed with a capacity of 1029 MWh and a maximum capacity of 300 MWh. Those number are unrealistic but in order to simulate the behavior of a \textit{price-taker} on the energy market we had to increase the battery capacity to create market power. We remind that batteries are \textit{price-maker} on smaller markets as ancillary services markets. The charging and discharging efficiency rate are set to 1 for simplicity.
When the battery behaves as a load, its bid is added directly to the total load of the grid. As a generator the battery bid is submitted to the clearing process described in Section \ref{model}.
\subsection{Results}
\begin{figure}[H]
\centering
\includegraphics[width=8.5cm]{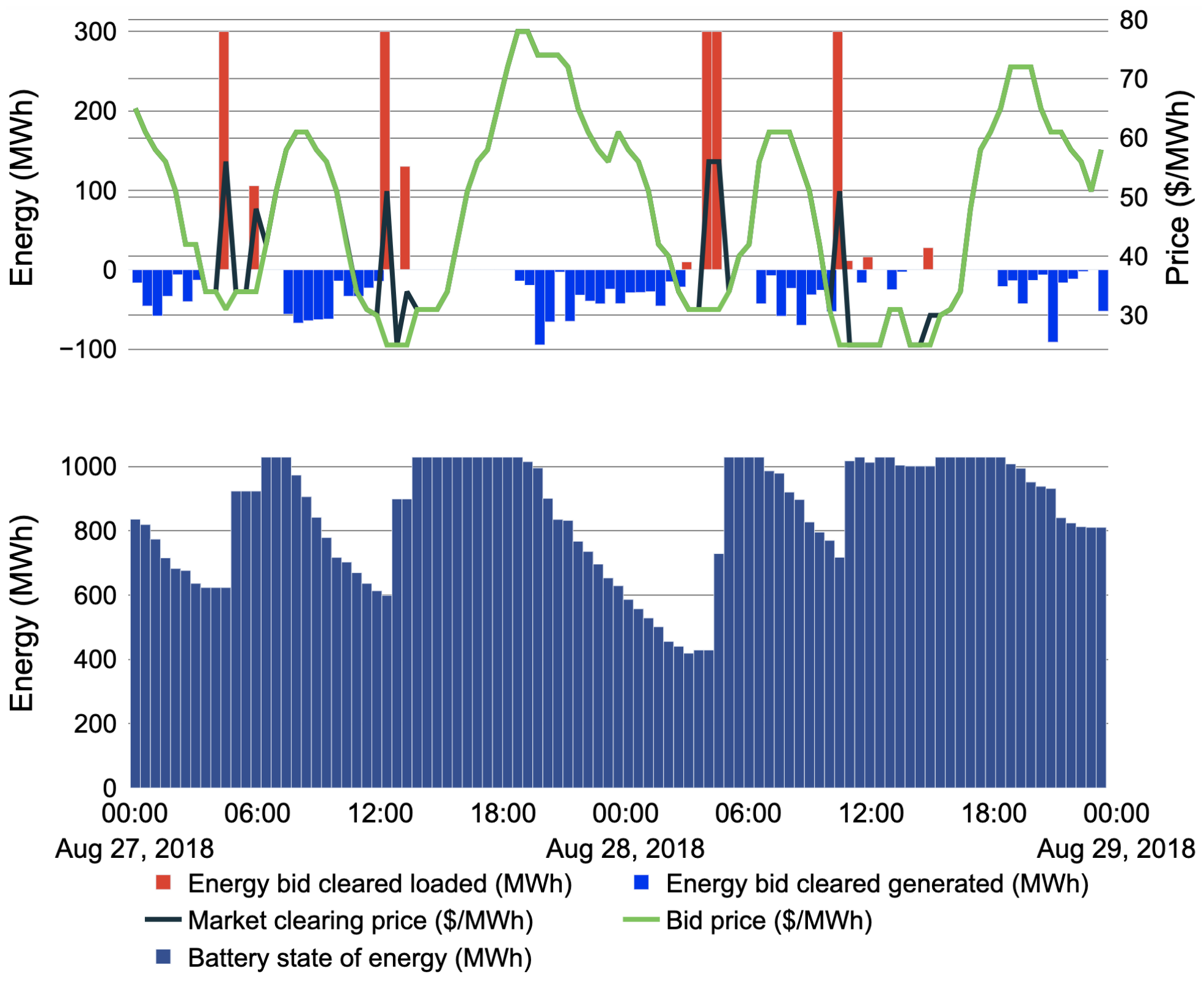}
\caption{MPC strategy over two days of the simulation}
\label{bids_energy_mpc}
\end{figure}

First, we evaluate that the MPC is behaving correctly, i.e. buying electricity at low price and selling at higher price. Figure \ref{bids_energy_mpc} plots the simulation result over two days. The bar plot in the above figure represents the generation and load bids capacity that were cleared by the market. The black line is the cleared price and the green line is the battery's bidding price. The MPC bids as a generator when prices are high and as a load when prices are low. An interesting observation is that the MPC algorithm bids at the clearing price when behaving as a generator. That is, the battery plays the role of the orange bid in Fig. \ref{clearing_process}. In this case, the market cannot clear the entire capacity bid since this is the marginal bid. As a consequence the battery is not discharging as much capacity as predicted by the MPC, this is the main source of loss for the MPC algorithm. Another observation is that when the battery behaves as a load, it increases the demand and consequently increases the prices due to the price-maker condition occurring.

In addition to showing the expected behavior, Fig. \ref{bids_energy_mpc} highlights the impact of the battery's bids on the market: Our system is a \textit{price-maker} on this market. 

\begin{figure}[H]
\centering
\includegraphics[width=9cm]{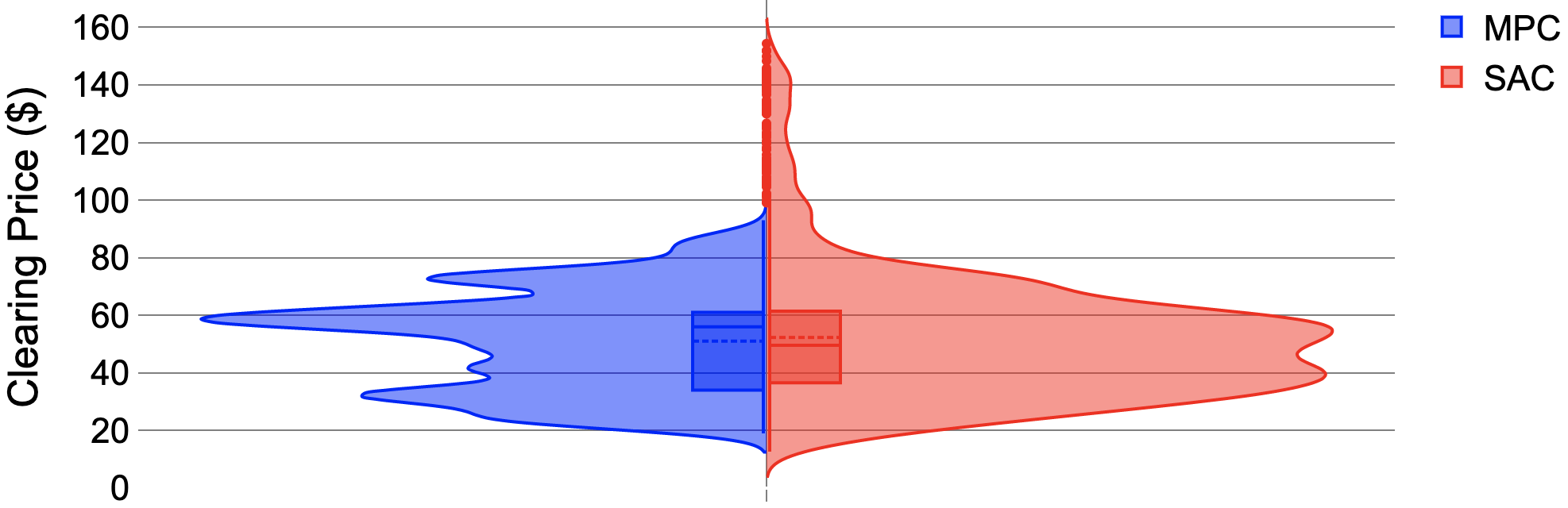}
\caption{Market Clearing Price Distribution for MPC and SAC}
\label{price_distr}
\end{figure}

\begin{figure}[H]
\centering
\includegraphics[width=8.8cm]{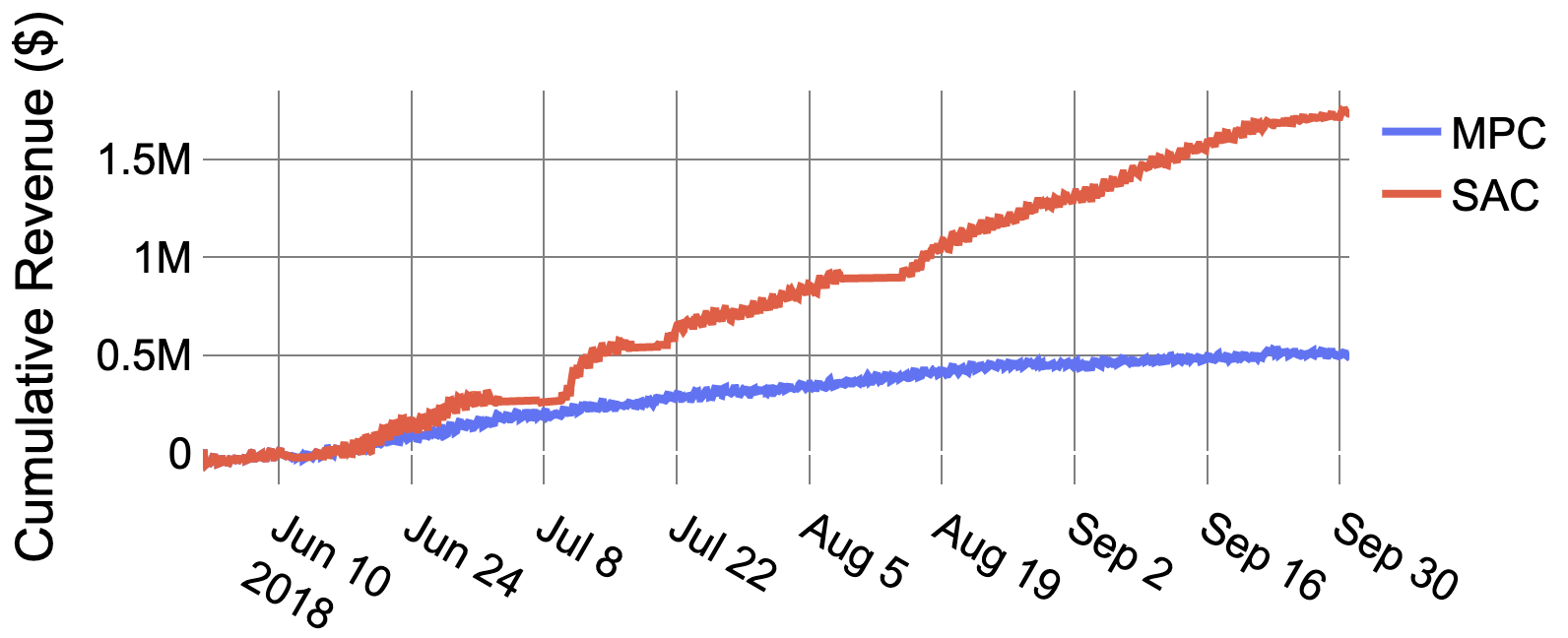}
\caption{Cumulative Revenue for MPC and SAC}
\label{revenue}
\end{figure}

We then apply the SAC algorithm to the same simulation. Our claim is that the SAC algorithm would recognize the battery's impact on the electricity market, and adapt. As a result, it is interesting but expected to see that the clearing price distribution varies depending on the bidding algorithm, as shown in Fig. \ref{price_distr}. 
To compare the SAC with the MPC algorithm, we compute in Table \ref{tab:bids} the total revenue earned by the two algorithms. The SAC significantly outperforms the MPC algorithm by earning over 3.5 times more revenue on average. In Fig.~\ref{revenue} we can see the cumulative revenue increasing at a much faster rate after the increase of $k$ over the time. 

The SAC proves to be a safe algorithm: only $6 \%$ of the bids are charging or discharging the battery above limits. Fortunately our system is built with a shield that blocs those actions.

\begin{figure}[H]
\centering
\includegraphics[width=8.8cm]{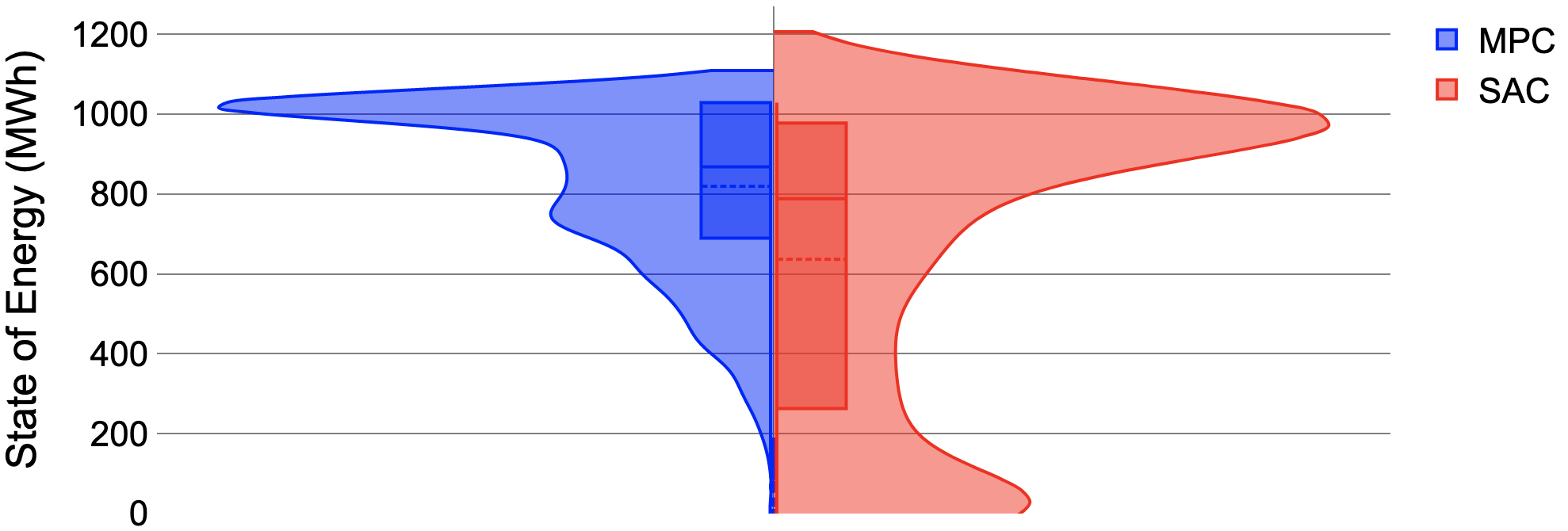}
\caption{Battery State of Energy Distribution for MPC and SAC}
\label{soe_distr}
\end{figure}

\begin{figure}[H]
\centering
\includegraphics[width=8.8cm]{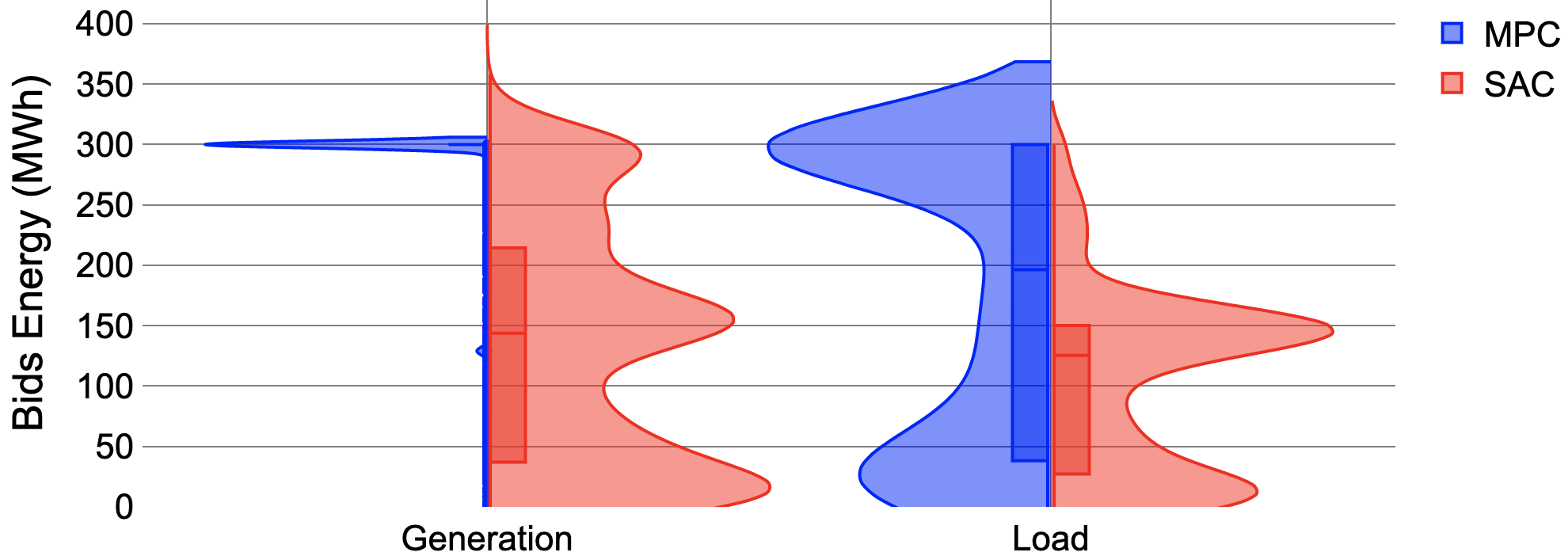}
\caption{Generation and Load Bids for MPC and SAC}
\label{bids_distr}
\end{figure}

To understand the behavioral differences between the two algorithms we plot the distribution of the battery state of energy in Fig. \ref{soe_distr}, as well as the distribution of capacity bids in Fig. \ref{bids_distr}.
The state of energy is varying across the entire range of possible values with SAC, taking full advantage of the battery's energy capacity. With MPC, the battery discharges less energy on the grid. This observation is related to the conclusion made on Fig. \ref{bids_energy_mpc}. Namely, the battery clears at the market clearing price and thus does not clear the entire capacity bid. We observe that the SAC algorithm solves this problem by recognizing the lower cleared quantity, and consequently adjusts the bid. This results in using the entire range of the battery capacity which leads to  a significantly higher revenue.
\begin{table}[!ht]
\renewcommand{\arraystretch}{1}
\caption{Bidding Strategies}
\label{tab:bids}
\centering
\begin{tabular}{c|c|c}
    \hline 
    & SAC  &  MPC\\
    \hline \hline
    Average revenue per day & \textbf{599 \$/day} & 175 \$/day \\
    \hline
    Average \% of bids capacity cleared & \textbf{80\%} & 15\% \\
    \hline
    Violations of battery capacity before shield &  6 \% & \textbf{0 \%} \\
    \hline
    \% of Generator bids & \textbf{66 \%} & 63 \% \\
    \hline
\end{tabular}
\end{table}

Yielding higher revenue charging and discharging cycles of the battery requires that the SAC generates more strategic bids. In Fig. \ref{bids_distr} we compare the distribution of capacity bids between the two algorithms. The MPC constantly bids as a generator at a capacity of 300 MWh, which is the maximum. This strategy is normal since the battery is most of the time at high state of energy and thus can discharge at maximum rate. The SAC strategy cannot bid more than the battery capacity allows, which yields a broad distribution  in Fig. \ref{soe_distr}. Another interesting behavior to point out is that the SAC generated load bids that are lower than MPC. However the load bids behave like an additional demand for the market and thus increases the clearing price a significant amount as seen in Fig. \ref{bids_energy_mpc}. Recall that the battery pays the market the clearing price when it operates as a load. The SAC recognizes that when the battery operates as a load, it inflates prices. Consequently, the SAC algorithm has lower capacity bids as to not inflate instantaneous costs.

\section{Conclusion}
Learning-based methods are particularly well suited to answer the need for optimal bidding algorithms. Indeed, markets have always been complex to model as they are multi-player and dynamic systems, but are necessary to understand when one player's bid impacts the clearing process, or when an ISO experiments with new market designs. For this reason, model-free or learning-based algorithms such as DRL are an appealing option. Those algorithms do not require a complete model of the system, and can adapt their actions in response to observed impacts on the system. DRL has proven its efficiency for controlling robots, playing video games, and targeting web content, but is rarely applied in energy systems. The main reason is because those algorithms learn through trial and error, and error can have dangerous consequences in real-world energy systems. Designing simulations to train these algorithms raises a paradox: if we can build a model of the environment, then why use a model-free algorithm? 

To use DRL in real-world energy system, we must overcome the danger of training the algorithm. Techniques like developing shields to limit dangerous actions, adding penalty terms to the reward function, or using supervisors to reduce the search in the action space have the potential to solve this issue. 

In this paper, we develop a Supervised Actor-Critic algorithm to optimally bid the energy of a \textit{price-maker} grid-scale battery on the electricity market. In addition, we use a shield as well as a penalty term in the reward to avoid dangerous actions.

The results show that those techniques improve bidding performance relative to the baseline \textit{price-taker} algorithm, while ensuring safety during algorithm training. 

This first approach is experimental, and developing those techniques at an industrial level would require more theoretical work on the action space reduction and the effect of a shield. Moreover, future work could study the reaction of the SAC to a market with dynamic and multiple \textit{price-maker} players.

\addtolength{\textheight}{-12cm}   






\bibliographystyle{IEEEtran}
\bibliography{IEEEabrv,main}

\end{document}